\documentclass[doublecol]{epl2}

\title{Continuous-time random walk theory of superslow diffusion}

\author{S. I. Denisov\inst{1,2}\thanks{E-mail: stdenis@pks.mpg.de}
\and H. Kantz\inst{1}}
\shortauthor{S. I. Denisov \etal}

\institute{
  \inst{1} Max-Planck-Institut f\"{u}r Physik komplexer Systeme -
  N\"{o}thnitzer Stra{\ss}e 38, D-01187 Dresden, Germany\\
  \inst{2} Sumy State University - Rimsky-Korsakov Street 2,
  UA-40007 Sumy, Ukraine
  }
  \pacs{05.40.Fb}{Random walks and L\'{e}vy flights}
  \pacs{02.50.Ey}{Stochastic processes}

\abstract{
Superslow diffusion, i.e., the long-time diffusion of particles whose
mean-square displacement (variance) grows slower than any power of time,
is studied in the framework of the decoupled continuous-time random walk
model. We show that this behavior of the variance occurs when the
complementary cumulative distribution function of waiting times is
asymptotically described by a slowly varying function. In this case,
we derive a general representation of the laws of superslow diffusion
for both biased and unbiased versions of the model and, to illustrate
the obtained results, consider two particular classes of waiting-time
distributions.}

\begin{document}

\maketitle

\section{Introduction}
Random processes in physical, biological, social, economic and other systems
often exhibit the anomalous diffusion behavior in the sense that at long times
the variance $\sigma^{2}(t)$ of these processes increases nonlinearly with time
\cite{BG, MK, AH}. Usually, the variance follows the power law, i.e.,
$\sigma^{2}(t) \propto t^{\alpha}$, and two types of anomalous diffusion,
subdiffusion (when $0<\alpha<1$) and superdiffusion (when $\alpha>1$), are
distinguished. The class of systems with such behavior of the variance is vast
and growing. Subdiffusion has been observed, e.g., in amorphous solids
\cite{SM}, percolation clusters \cite{KMK} and living cells \cite{GC,SW}, and
superdiffusion in turbulent flows \cite{Rich,SWS}, optical lattices \cite{KSW}
and foraging animals \cite{Sim,Hum}.

In general, the time-dependence of the variance is not restricted by a power
function. For example, under certain conditions the power law can be modified
by a logarithmic function of time (see, e.g., refs.~\cite{Shl,ZK,DK}). But
since the logarithmic function varies much more slowly than the power one, this
modification is not so important. In contrast, in the case of
\textit{superslow} diffusion, which is defined by the condition that
$\sigma^{2}(t)/ t^{\alpha} \to 0$ as $t \to \infty$ for all $\alpha>0$, the
variance does not follow the power law at long times. The Sinai diffusion for
which $\sigma^{2}(t) \propto \ln^{4} t$ \cite{Sin} represents the best known
example of this type of diffusion. Some other examples of superslow diffusion
have been found in resistor networks \cite{HBSSB}, continuous-time random walks
(CTRWs) \cite{HW}, charged polymers \cite{SSB}, aperiodic environments
\cite{ITR}, iterated maps \cite{DrKl}, Langevin dynamics \cite{DH}, fractional
kinetics \cite{CKS}, etc. A common feature of all these examples is that the
long-time dependence of the variance, i.e., the law of superslow diffusion, is
given by a power function of the logarithm of time: $\sigma^{2}(t) \propto
\ln^{\nu} t \;(\nu>0)$.

A natural question is if there exist laws of superslow diffusion that differ
from $\sigma^{2}(t) \propto \ln^{\nu} t$. Since the defining condition
$\sigma^{2}(t)/ t^{\alpha} \to 0$ holds not only for this particular law, one
can expect that other laws also exist. In this letter we use the decoupled CTRW
model to show that if the waiting-time distributions \textit{vary slowly} at
infinity (see below), then the corresponding laws of superslow diffusion form a
wide class of slowly varying functions that increase without limit as time
evolves.

\section{Description of the model}
Within the CTRW formalism (see, e.g., refs.~\cite{BG, MK, AH}), the position
$X(t)$ ($X(0)=0$) of a diffusing particle is described by the continuous-time
jump process
\begin{equation}
    X(t) = \sum_{n=1}^{N(t)} x_{n},
\label{X(t)}
\end{equation}
where $N(t) \in (0,1,2,\ldots)$ is the random number of jumps up to time $t$
(if $N(t)=0$ then $X(t)=0$) and $x_{n} \in (-\infty, \infty)$ is the random
magnitude of the $n$-th jump. The number of jumps $N(t)$ is characterized by
the waiting times $\tau_{n} \in [0,\infty)$, i.e., random times between the
particle jumps. These times and the jump magnitudes $x_{n}$ are assumed to be
independent and identically distributed with the probability densities
$p(\tau)$ and $w(x)$, respectively. In the decoupled CTRW model it is also
assumed that two sets of variables $\tau_{n}$ and $x_{n}$ are independent of
each other, so that the joint probability density of these variables is given
by the product $p(\tau)w(x)$.

In Fourier-Laplace space, the probability density $P(x,t)$ of the process
$X(t)$ satisfies, in  the decoupled case, the Montroll-Weiss equation \cite{MW}
\begin{equation}
    P_{ks} = \frac{1-p_{s}}{s(1-p_{s}w_{k})},
\label{M-Weq}
\end{equation}
where the Fourier and Laplace transforms are defined as $u_{k} = \mathcal{F}
\{u(x)\} = \int_{-\infty}^{\infty} dx e^{ikx} u(x)$ and $v_{s} = \mathcal{L}
\{v(t)\} = \int_{0}^{\infty} dt e^{-st} v(t)\; (\mathrm{Re}\, s>0)$,
respectively. Using eq.~(\ref{M-Weq}), the Laplace transform of the $m$-th
($m=1,2,\ldots$) moment of $X(t)$ can be written in the form
\begin{equation}
    \langle X^{m}(t) \rangle_{s} = (-i)^{m} \frac{1-p_{s}}{s}
    \frac{d^{m}}{dk^{m}} \frac{1}{1-p_{s}w_{k}} \bigg|_{k=0}
\label{Xm_s}
\end{equation}
with the angular brackets denoting an average over the sample paths of the
random process $X(t)$. Applying to eq.~(\ref{Xm_s}) the inverse Laplace
transform, $v(t) = \mathcal{L}^ {-1} \{v_{s}\} = (1/2\pi i) \int_{c-
i\infty}^{c+ i\infty} ds e^{st} v_{s}$ (it is assumed that the real parameter
$c$ exceeds the real parts of all singularities of $v_{s}$), one obtains
$\langle X^{m}(t) \rangle = \mathcal{L} ^{-1} \{ \langle X^{m}(t) \rangle_{s}
\}$.

In this work we focus on the long-time behavior of the variance of the particle
position, defined as
\begin{equation}
    \sigma^{2}(t) = \langle X^{2}(t) \rangle - \langle X(t) \rangle^{2},
\label{var}
\end{equation}
under the additional condition that \textit{all} fractional moments of the
waiting-time distribution are infinite, i.e., $\mathcal{T}_{\rho} =
\int_{0}^{\infty} d\tau \tau^{\rho} p(\tau) = \infty$ at $\rho>0$. By analogy
with \cite{DKH}, we call the probability densities $p(\tau)$ satisfying this
condition the \textit{super-heavy-tailed} densities. It should be noted that a
special class of these densities with $p(\tau) \sim a/(\tau \ln^{1+\nu}\tau)$
$(a>0,\, \tau \to \infty)$ has already been put forward by Havlin and Weiss
\cite{HW}. But here we consider a much wider class of super-heavy-tailed
probability densities that are characterized by the asymptotic behavior
\begin{equation}
    p(\tau) \sim \frac{h(\tau)}{\tau} \quad (\tau \to \infty),
\label{p as}
\end{equation}
where a positive function $h(\tau)$ is slowly varying at infinity in the sense
that the condition $h(\mu \tau) \sim h(\tau)$ holds for all $\mu>0$ as $\tau
\to \infty$. Since the waiting-time distributions are assumed to be normalized,
$\mathcal{T}_{0} = \int_{0}^{\infty}d\tau p(\tau) = 1$, the admissible
functions $h(\tau)$ form a subclass of slowly varying functions that at $\tau
\to \infty$ tend to zero fast enough.

\section{Laws of superslow diffusion}
In order to ensure the finiteness of the variance $\sigma^{2}(t)$, we assume
that the first two moments of $w(x)$, $l_{1} = \int_{-\infty}^ {\infty}
dxxw(x)$ and $l_{2} = \int_{-\infty}^ {\infty} dxx^{2}w(x)$, exist. In this
case, using the definition of the Fourier transform of $w(x)$, one obtains
\begin{equation}
    w_{k} = 1 + il_{1}k - \frac{1}{2}\, l_{2}k^{2} + o(k^{2})
    \quad (k \to 0),
\label{wk}
\end{equation}
and so eq.~(\ref{Xm_s}) for $m=1$ and $m=2$ gives
\begin{equation}
    \langle X(t) \rangle_{s} = \frac{l_{1}p_{s}}{s(1-p_{s})}, \quad
    \langle X^{2}(t) \rangle_{s} = \frac{2l_{1}^{2}p_{s}^{2}}
    {s(1-p_{s})^{2}} +  \frac{l_{2}p_{s}}{s(1-p_{s})}.
\label{X1,2s1}
\end{equation}
Our next step is to find the asymptotic expressions for $\langle X(t)
\rangle_{s}$ and $\langle X^{2}(t) \rangle_{s}$ when a \textit{positive}
parameter $s$ approaches zero. The motivation is that if these expressions
satisfy certain conditions then the long-time behavior of the moments $\langle
X(t) \rangle$ and $\langle X^{2}(t) \rangle$ can be found directly from the
Tauberian theorem for the Laplace transform.

Let us first represent the Laplace transform of $p(\tau)$ as
\begin{equation}
    p_{s} = 1 - \int_{0}^{\infty}d\tau (1-e^{-s\tau}) p(\tau).
\label{ps1}
\end{equation}
Then, introducing the probability
\begin{equation}
    V(t) = \int_{t}^{\infty}d\tau p(\tau)
\label{V}
\end{equation}
that the waiting time exceeds $t$ and using the new variable of integration $q=
s\tau$, we can reduce eq.~(\ref{ps1}) to the form
\begin{equation}
    p_{s} = 1 - \int_{0}^{\infty}dq e^{-q} V\! \left( \frac{q}{s}
    \right).
\label{ps2}
\end{equation}
The function $V(t)$, known as the complementary cumulative distribution
function of waiting times (or exceedance probability), satisfies the initial
condition $V(0)=1$ and tends to zero in a slowly varying way as $t\to \infty$.
The last property of $V(t)$, which is crucial for our approach, can be
straightforwardly verified using the asymptotic formula (\ref{p as}) and the
condition $h(\mu \tau) \sim h(\tau)$ ($\tau \to \infty$):
\begin{equation}
    V (\mu t) \sim \int_{\mu t}^{\infty}d\tau \frac{h(\tau)}{\tau}
    \sim \int_{t}^{\infty}dz \frac{h(z)}{z} \sim V(t)
\label{rel1}
\end{equation}
($t \to \infty$). We note in this connection that the waiting-time cumulative
distribution function $F(t) = \int_{0}^{t} d\tau p(\tau) = 1- V(t)$ also varies
slowly at infinity.

Since $V(\mu t) \sim V(t)$ as $t \to \infty$, the function $V(q/s)$ in
eq.~(\ref{ps2}) at $s \to 0$ can be replaced by $V(1/s)$, yielding
\begin{equation}
    p_{s} \sim 1 - V\! \left( \frac{1}{s} \right) \quad (s \to 0).
\label{ps3}
\end{equation}
Utilizing this result in eq.~(\ref{X1,2s1}), we find, in the limit of small
$s$, that
\begin{equation}
    \langle X(t) \rangle_{s} \sim \frac{l_{1}}{sV(1/s)}, \quad
    \langle X^{2}(t) \rangle_{s} \sim \frac{2l_{1}^{2}}
    {sV^{2}(1/s)}
\label{X1,2s2}
\end{equation}
if $l_{1} \neq 0$ and
\begin{equation}
    \langle X(t) \rangle_{s} =0, \quad
    \langle X^{2}(t) \rangle_{s} \sim  \frac{l_{2}}{sV(1/s)}
\label{X1,2s3}
\end{equation}
if $l_{1}=0$. Next, to find the asymptotic behavior of the moments $\langle
X(t) \rangle$ and $\langle X^{2}(t) \rangle$, we apply the Tauberian theorem
for the Laplace transform, which states (see, e.g., ref.~\cite{Fel}) that if
the function $v(t)$ is ultimately monotone and $v_{s} \sim s^{-\gamma} L(1/s)$
as $s\to 0$ then $v(t) \sim t^{\gamma-1} L(t)/\Gamma(\gamma)$ as $t\to \infty$.
Here, $L(t)$ is a slowly varying function at infinity, $\Gamma(\gamma)$ is the
gamma function, and $0<\gamma<\infty$. In accordance with this theorem,
eqs.~(\ref{X1,2s2}) and (\ref{X1,2s3}) lead to
\begin{equation}
    \langle X(t) \rangle \sim \frac{l_{1}}{V(t)}, \quad
    \langle X^{2}(t) \rangle \sim \frac{2l_{1}^{2}} {V^{2}(t)}
\label{X1,2a}
\end{equation}
and
\begin{equation}
    \langle X(t) \rangle =0, \quad
    \langle X^{2}(t) \rangle \sim  \frac{l_{2}}{V(t)},
\label{X1,2b}
\end{equation}
respectively. We emphasize that in the biased case, when $l_{1} \neq 0$, the
main terms of the asymptotic expansion of $\langle X(t) \rangle^{2}$ and
$\langle X^{2}(t) \rangle$ are different, see eq.~(\ref{X1,2a}). The same
situation occurs also for the biased subdiffusion, while for the biased normal
diffusion $\langle X^{2}(t) \rangle \sim \langle X(t) \rangle^{2}$ and $\langle
X^{2}(t)\rangle - \langle X(t) \rangle^{2} \propto t$ \cite{Shl, DDK}. These
features of the moments arise from the fact that, since in the biased case
$P_{-ks} \neq P_{ks}$, the left and right tails of the probability density
$P(x,t)$ are different.

Thus, using eqs.~(\ref{var}), (\ref{X1,2a}) and (\ref{X1,2b}), for the
long-time behavior of the variance one immediately obtains
\begin{equation}
    \sigma^{2}(t) \sim \left\{\!\! \begin{array}{ll}
    l^{2}_{1}/V^{2}(t),
    & l_{1} \neq 0
    \\ [6pt]
    l_{2}/V(t),
    & l_{1} = 0
    \end{array}
    \right.
\label{var1}
\end{equation}
with
\begin{equation}
    V(t) \sim \int_{t}^{\infty}d\tau \frac{1}{\tau} h(\tau).
\label{V1}
\end{equation}
The asymptotic formula (\ref{var1}), which describes both biased ($l_{1} \neq
0$) and unbiased ($l_{1}=0$) diffusion in systems with super-heavy-tailed
distributions of waiting times, is the central result of this letter. The most
important is the fact that the exceedance probability $V(t)$ varies slowly at
infinity. It is this feature of $V(t)$ that is responsible for the superslow
character of diffusion in these systems. Indeed, it is well known \cite{BGT}
that if some function, say $V(t)$, varies slowly, so does any power of this
function. Hence, the function $\sigma^{2}(t)$ is also slowly varying, and thus
the condition $\sigma^{2}(t)/t^{\alpha} \to 0$ holds for all $\alpha>0$ as $t
\to \infty$ \cite{BGT}. In other words, the variance $\sigma^{2}(t)$ grows to
infinity ($\sigma^{2}(t) \to \infty$ because $V(t) \to 0$ as $t \to \infty$)
slower than any positive power of $t$, i.e., superslow diffusion occurs. It is
worth noting that the slower the exceedance probability decreases, the slower
this diffusion. At the same time, since $\sigma^{2}(t) |_{l_{1} \neq 0} /
\sigma^{2}(t) |_{l_{1} = 0} \to \infty$ as $t\to \infty$, the biased superslow
diffusion is much faster than the unbiased one.

Because the exceedance probability $V(t)$ is slowly varying, the Karamata
representation theorem \cite{BGT} suggests that $V(t)$ at $t \to \infty$ can
always be written in the form
\begin{equation}
    V(t) \sim v \exp\! \left( -\int_{\kappa}^{t}d\tau
    \frac{1}{\tau} \epsilon(\tau) \right)\! .
\label{V2}
\end{equation}
Here, $v$ is a positive parameter and $\kappa$ is a non-negative parameter
which, without loss of generality, can be taken to be zero. In general, for an
arbitrary slowly varying function $V(t)$ it is only required that the function
$\epsilon(\tau)$ tends to zero as $\tau \to \infty$. But in our case $V(t) \to
0$ in the long-time limit. Therefore, the function $\epsilon(\tau)$ must be
positive and satisfy the conditions $\epsilon(\tau) \to 0$ and $\int_{\kappa}
^{\tau}d\tau' \epsilon(\tau')/\tau' \to \infty$ as $\tau \to \infty$. The
asymptotic formula (\ref{var1}) with so defined exceedance probability $V(t)$
represents the most general form of the diffusion laws in the reference model.
If the asymptotic behavior of the function $h(\tau)$ is known, then the
asymptotic behavior of the function $\epsilon(\tau)$ and the magnitude of the
parameter $v$ can be determined as
\begin{equation}
    \epsilon(\tau) \sim \frac{h(\tau)}{V(\tau)} \quad (\tau \to \infty)
\label{eps}
\end{equation}
and
\begin{equation}
    v = \lim_{\tau \to \infty} \exp\! \left( \int_{\kappa}^{\tau}
    d\tau' \frac{1}{\tau'} \epsilon(\tau') \right) \! V(\tau)
\label{v}
\end{equation}
with $V(\tau)$ taken from eq.~(\ref{V1}). It should be emphasized that because
the functions $h(\tau)$ and $V(\tau)$ are slowly varying, their ratio, the
function $\epsilon(\tau)$, also varies slowly. This property holds for all
waiting-time probability densities characterized by the asymptotic behavior
(\ref{p as}).

We complete this section by noting that the laws of superslow diffusion
(\ref{var1}) are obtained within the simplest CTRW model. The fact that these
laws include the diffusion law $\sigma^{2}(t) \propto \ln^{\nu} t$ as a very
special case does not mean, of course, that this model describes in detail the
systems mentioned in the Introduction. In this context we recall that our goal
is not the application of the decoupled CTRW model to those systems but rather
the use of this model to show that the laws of superslow diffusion constitute a
broad class of slowly varying functions.

\section{Illustrative examples}
As a first example of super-heavy-tailed distributions of waiting times whose
asymptotic behavior is described by eq.~(\ref{p as}), we consider a
two-parametric class of such distributions characterized by the probability
density
\begin{equation}
    p(\tau) = \frac{(r-1)\ln^{r-1}\eta}{(\eta + \tau)
    \ln^{r} (\eta + \tau)}
\label{p1}
\end{equation}
with the parameters $r>1$ and $\eta>1$ (these conditions guarantee that
$p(\tau)$ is positive and normalized). In this case the exceedance probability
(\ref{V}) is calculated exactly,
\begin{equation}
    V(t) = \bigg(\frac{\ln \eta}{\ln(\eta+t)}\bigg)^{r-1},
\label{V3}
\end{equation}
and thus $V(t) \sim (\ln\eta / \ln t)^{r-1}$ as $t\to \infty$. The substitution
of the last asymptotic result into eq.~(\ref{var1}) leads to the following
asymptotic formula:
\begin{equation}
    \sigma^{2}(t) \sim \left\{\!\! \begin{array}{ll}
    (l_{1}/\ln^{r-1}\eta)^{2}\ln^{2(r-1)}t,
    & l_{1} \neq 0
    \\ [6pt]
    (l_{2}/\ln^{r-1}\eta)\ln^{r-1}t,
    & l_{1} = 0,
    \end{array}
    \right.
\label{var2}
\end{equation}
which for $l_{1} = 0$ was first obtained in ref.~\cite{HW}. Finally, since
according to eq.~(\ref{p1}) $h(\tau) \sim (r-1)\ln^{r-1}\eta/ \ln^{r} \tau$ as
$\tau \to \infty$, from eqs.~(\ref{eps}) and (\ref{v}) we find $\epsilon (\tau)
\sim (r-1)/\ln \tau$ and $v = \ln^{r-1} \eta$.

Our second example is characterized by the waiting-time probability density of
the form
\begin{equation}
    p(\tau) = \frac{(r-1)(\ln\ln \eta)^{r-1}}{(\eta + \tau)
    \ln(\eta + \tau) [\ln\ln (\eta + \tau)]^{r}}
\label{p2}
\end{equation}
($r>1$, $\eta>e$). The main feature of this density is that its right tail is
heavier than in the previous case, i.e., $p(\tau)$ at $\tau \to \infty$ tends
to zero slower than the probability density from eq.~(\ref{p1}). Therefore, the
exceedance probability
\begin{equation}
    V(t) = \bigg(\frac{\ln\ln \eta}{\ln\ln(\eta+t)}\bigg)^{r-1},
\label{V4}
\end{equation}
which corresponds to the probability density (\ref{p2}), decreases slower than
$V(t)$ from eq.~(\ref{V3}) and, as a consequence, the variance (\ref{var1})
increases with time slower than in the previous case:
\begin{equation}
    \sigma^{2}(t) \sim \left\{\!\! \begin{array}{ll}
    [l_{1}/(\ln\ln \eta)^{r-1}]^{2}(\ln\ln t)^{2(r-1)},
    & l_{1} \neq 0
    \\ [6pt]
    [l_{2}/(\ln\ln \eta)^{r-1}](\ln\ln t)^{r-1},
    & l_{1} = 0.
    \end{array}
    \right.
\label{var3}
\end{equation}
Taking also into account that according to eq.~(\ref{p2}) $h(\tau) \sim
(r-1)(\ln\ln \eta)^{r-1} /[\ln\tau (\ln\ln \tau)^{r}]$ as $\tau \to \infty$,
for this case we obtain $\epsilon (\tau) \sim (r-1)/(\ln \tau \ln\ln \tau)$ and
$v = (\ln\ln \eta)^{r-1}$.

The above examples are illustrative and, of course, the diffusion laws
(\ref{var2}) and (\ref{var3}) do not exhaust all possible laws of superslow
diffusion. As is shown in this work, the possible laws form a class of slowly
varying at infinity functions $\sigma^{2}(t)$ satisfying the condition
$\sigma^{2}(t) \to \infty$ as $t \to \infty$. This class is sufficiently large
because it contains infinitely many functions which at $t\to \infty$ grow both
faster and slower than a given function from this class. For example, the
function $\sigma^{2}(t) \propto \exp(\ln^{\beta} t)/\ln t$ ($0<\beta<1$) grows
faster than $\sigma^{2}(t)$ from eq.~(\ref{var2}), and the function
$\sigma^{2}(t) \propto \ln \ln \ln t$ grows slower than $\sigma^{2}(t)$ from
eq.~(\ref{var3}). It should also be stressed that, although for the probability
densities $p(\tau)$ characterized by the condition (\ref{p as}) the mean
waiting time is infinite, the regime of superslow diffusion is reached at
finite times exceeding the longest time scale of the corresponding density. In
particular, the diffusion laws (\ref{var2}) and (\ref{var3}) take place at $t
\gg \eta$.

\section{Conclusions}
Using the decoupled continuous-time random walk model characterized by the
waiting-time distribution functions that vary slowly at infinity, we have
determined a wide class of the superslow diffusion laws. This class consists of
the slowly varying functions, describing the variance of the particle position,
which tend to infinity slower than any power of time. We have established the
connection between the long-time behavior of the variance and the asymptotic
behavior of the exceedance probability. Specifically, in the cases of biased
and unbiased superslow diffusion the variance is inversely proportional to the
second and first powers of the exceedance probability, respectively. Due to
this difference, the biased superslow diffusion at long times is much faster
than the unbiased one. Finally, by applying the Karamata representation
theorem, we have found the most general form of the laws of superslow diffusion
in the reference model.

\end{document}